\documentclass{aip-cp}

\usepackage[numbers]{natbib}
\usepackage{rotating}
\usepackage{graphicx}

\begin{document}

\title{A New Eye on the VHE Transient Universe with the HAWC Online Flare Monitor}

\author[aff1]{Thomas Weisgarber\corref{cor1}}
\author{the HAWC Collaboration\corref{cor2}}

\affil[aff1]{Wisconsin IceCube Particle Astrophysics Center (WIPAC) and Department of Physics, University of Wisconsin--Madison, Madison, WI, USA}
\corresp[cor1]{Corresponding author: weisgarber@wisc.edu}
\corresp[cor2]{For a complete author list, see http://hawc-observatory.org/collaboration}

\maketitle

\begin{abstract}
The High Altitude Water Cherenkov (HAWC) Observatory recently began full-scale operations, surveying 2/3 of the entire sky at very high energy (VHE; $E>100$ GeV).
This new view of the sky offers the opportunity to detect flares from blazars, facilitating studies of the mechanisms powering their central engines and providing an avenue to constrain the properties of particles and fields in intergalactic space.
The HAWC Collaboration has implemented an online flare monitor to search for rapid and extreme transient activity from a set of blazars either known or suspected to produce VHE emission.
The goal of this project is to issue alerts sufficiently rapidly to form a complete multiwavelength picture of the flare.
We describe the current status of the online flare monitor, demonstrating its ability to detect flares via a study of the blazars Markarian 421 and Markarian 501 in offline data.

\end{abstract}

\section{Blazar Flares at Very High Energies}

Blazars comprise a class of active galactic nuclei (AGNs) with relativistic jets oriented along the line of sight to Earth~\citep{Urry:1995ky}.
The broadband spectra of blazars characteristically exhibit a ``two-bump'' feature in $\nu F_\nu$ space, commonly interpreted as a low-energy bump arising from synchrotron radiation and a high-energy bump from inverse Compton scattering in the leptonic-production scenario or from hadronic processes such as pion decay in the hadronic-production scenario (see, for example,~\citep{Bottcher:2013hr} and references therein).
These spectra extend to very high energy (VHE; $E>100$ GeV) and exhibit variability on time scales ranging from minutes to months with flux increases of more than an order of magnitude~\citep{Tluczykont:2007bt}.
Over 60 blazars have been detected in the VHE band, with the vast majority belonging to the high-frequency peaked BL Lac (HBL) class~\citep{tevcat}.

The identification and multiwavelength study of blazar flaring states offers insight into the mechanisms powering relativistic jets in AGNs.
In particular, the particle acceleration mechanisms and dominant contributing particle types (leptonic or hadronic) may be constrained by the observation of flares, during which acceleration is believed to take place.
Furthermore, the increase in high energy event counts during flares supports a number of secondary studies.
Included among these are measurements of the extragalactic background light (EBL)~\citep{Mazin:2013ky}, searching for time delays caused by an intergalactic magnetic field (IGMF)~\citep{Neronov:2009fo,Weisgarber:ICRC2013}, and searching for possible effects of Lorentz invariance violation (LIV)~\citep{Nellen:ICRC2015}.
To facilitate these studies, it is important to identify flares rapidly so that a complete multiwavelength picture of the flares may be obtained via coordinated observations from many instruments.
Of particular interest are very extreme events such as the July 2006 flare of PKS 2155-304, which lasted for $\sim1$ hour.
During this time, the blazar produced VHE fluxes in excess of 10 times the flux from the Crab Nebula above 200 GeV and showed variability on the scale of minutes~\citep{Aharonian:2007ep}.

Due to their excellent sensitivity, imaging atmospheric Cherenkov telescopes (IACTs) are optimal for observing flares from blazars.
However, their limited field of view and $\sim15$\% duty cycle render them non-ideal for searching for VHE flares, since only one source may be observed at a time.
The High Altitude Water Cherenkov (HAWC) Observatory offers a complementary view of the sky, operating with a wide field of view of $\sim2$ sr and a duty cycle of $>95$\%.
Although less sensitive than the IACTs, HAWC is currently the most sensitive VHE survey instrument ever constructed.
HAWC provides an unbiased view of all blazars crossing its field of view, enabling it to send alerts when transient phenomena are detected.


\section{The HAWC Observatory}

The HAWC Observatory is located at $19^\circ$ N latitude at an elevation of 4100 m in the state of Puebla, Mexico.
HAWC comprises 300 optically isolated water Cherenkov detectors (WCDs).
Each WCD is instrumented with 4 upward-facing photomultiplier tubes (PMTs), which are used to collect the light from charged particles in extensive air showers initiated by gamma rays and cosmic rays in the energy range from $\sim100$ GeV to $>100$ TeV.
The charge and timing information from each PMT is used to reconstruct the energy and arrival direction of the primary particle.
Completed on 20 March 2015, HAWC conducts unbiased observations of the sky between $-26^\circ$ and $64^\circ$ in declination, surveying 2/3 of the sky every day.
Further details on the operation of HAWC are presented in~\citep{Sandoval:Gamma2016}.

\section{Performance of the HAWC Online Flare Monitor}

The HAWC Collaboration has implemented several searches for VHE transients~\citep{Lauer:Gamma2016}.
Among these, the online flare monitor searches for flares with fluxes reaching the multi-Crab range on sub-day timescales.
The flare monitor employs the Bayesian Block algorithm~\citep{Scargle:2013dm} to determine whether the observations from a predetermined list of sources are consistent with a constant flux.
We presently monitor 186 sources within the sky coverage of HAWC, including all known VHE blazars and a selection of nearby blazars ($z<1$) from the 2FHL catalog~\citep{Ackermann:2016gt}.

\begin{figure}[t!]
\centerline{\includegraphics[width=0.6\textwidth]{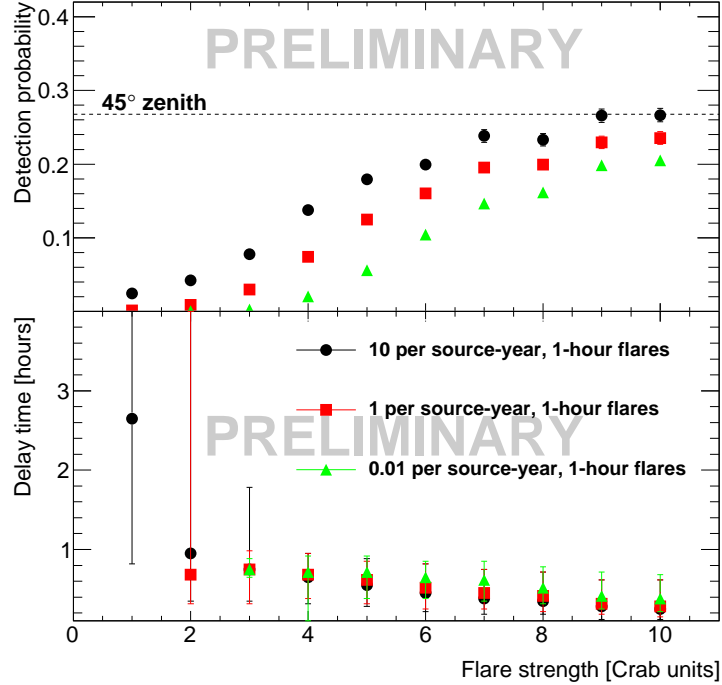}}
\caption{Sensitivity of the flare monitor to flares of varying strengths for sources located at the declination of the Crab Nebula. The top panel shows the probability of detection, which converges to the dashed line showing the fraction of time that the source spends within the HAWC field of view. In the bottom panel, the points depict the median time to flare detection, while the error bars show the central 68\% of the distribution. The different points show the results for false alarm rates of 10, 1, and 0.01 triggers per source per year.}
\label{fig:performance}
\end{figure}

The performance of the HAWC online flare monitor is summarized in Figure~\ref{fig:performance}, which shows the probability of detection and time to detection as a function of flare strength.
Flares scaled to the measured flux from the Crab Nebula with a duration of 1 hour were injected randomly into the data.
We chose to inject flares completely uniformly in time, even when the source is not in the field of view, in order to avoid the problem of defining the {\it detectability} of flares occurring at large zenith angles.
As a result, the probability of detection for strong flares does not converge to 1, but rather toward the fraction of time that the source spends in the HAWC field of view.
As is apparent from Figure~\ref{fig:performance}, the online flare monitor can detect flares with strengths equal to a few to several times the flux from the Crab Nebula, and it does so sufficiently rapidly to alert other instruments before the flare has ended.

\begin{figure}
\centerline{\includegraphics[width=\textwidth]{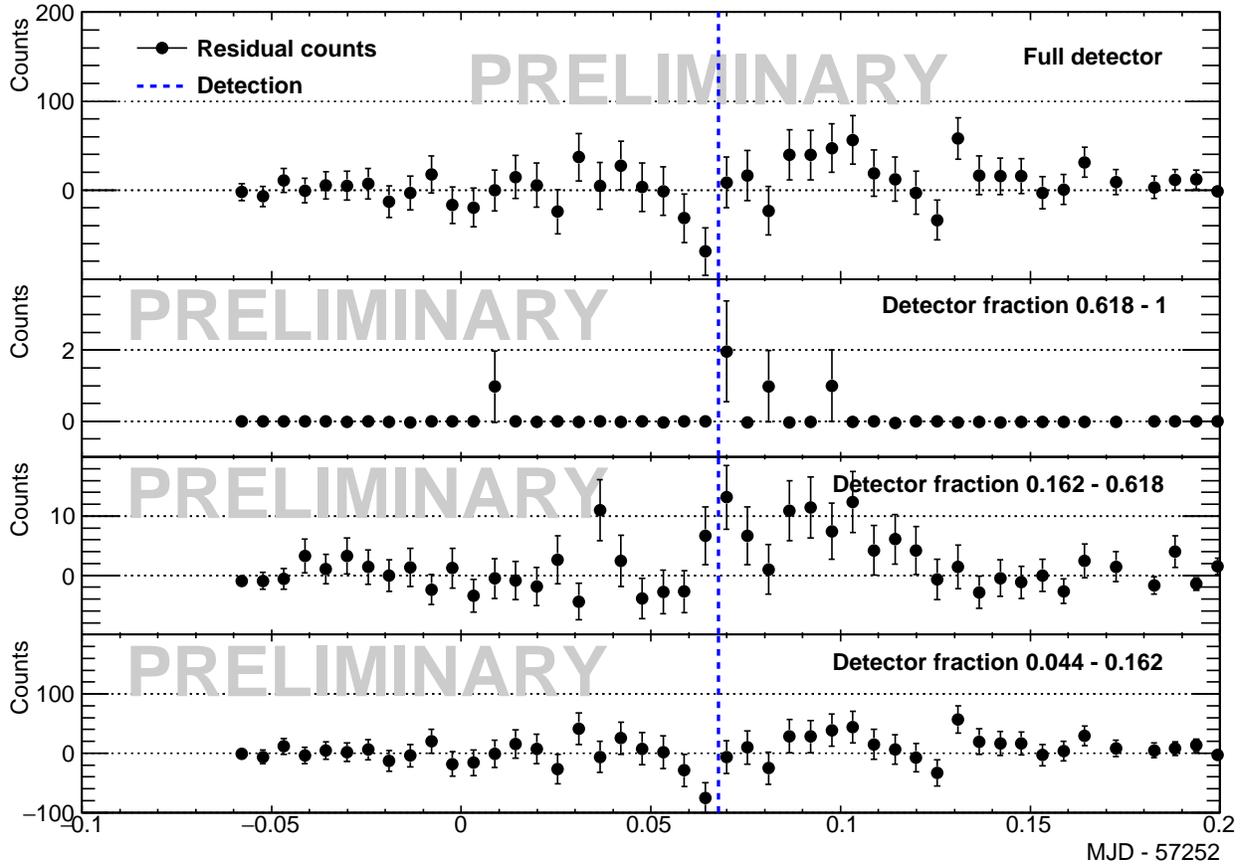}}
\caption{Excess counts from the flare monitor showing the detection of a flare from the blazar Markarian 501. The top panel shows the full detector excess counts, while the bottom 3 panels show the excess counts in bins corresponding to different fractions of the number of PMTs in the detector contributing to the events. The vertical blue dashed line represents the start time of the flare as identified by the flare monitor.}
\label{fig:flare}
\end{figure}

We further verify the performance of the online flare monitor by applying it to offline data collected between 26 November 2014 and 14 December 2015.
We limit our search to the 2 nearest known VHE blazars, Markarian 421 and Markarian 501.
An example of a flare detected by the online flare monitor during this period appears in Figure~\ref{fig:flare}.
This flare was detected from the blazar Markarian 501.
It occurred at 01:37 UTC on 18 August 2015 and lasted for approximately 1 hour.
The flare monitor associates this flare with an equivalent false alarm rate of less than 0.01 events per source per year, meaning that we would expect to see a background fluctuation of this magnitude less than once every 100 years from the source.
Analysis of the properties of this flare is ongoing, and a search of the offline data for additional flares from other sources is currently in progress.

\section{Conclusion}

At present, the HAWC online flare monitor is operating with sensitivity to extremely strong and rapid flares.
The capability of the flare monitor has been demonstrated via detection of a flare from Markarian 501 in offline data.
Monitoring of the online data is currently in progress, and we expect to provide alerts to the general community soon.

\section{Acknowledgments}
We acknowledge the support from: the US National Science Foundation (NSF); the US Department of Energy Office of High-Energy Physics; the Laboratory Directed Research and Development (LDRD) program of Los Alamos National Laboratory; Consejo Nacional de Ciencia y Tecnolog\'{\i}a (CONACyT), Mexico (grants 271051, 232656, 167281, 260378, 179588, 239762, 254964, 271737); Red HAWC, Mexico; DGAPA-UNAM (grants RG100414, IN111315, IN111716-3, IA102715, 109916); VIEP-BUAP; the University of Wisconsin Alumni Research Foundation; the Institute of Geophysics, Planetary Physics, and Signatures at Los Alamos National Laboratory; Polish Science Centre grant DEC-2014/13/B/ST9/945.


\bibliographystyle{text-arxiv}%
\bibliography{text-arxiv}%

\end{document}